\documentclass[12pt,showpacs,preprintnumbers,amsmath,amssymb]{revtex4}

\usepackage{graphicx}
\usepackage{dcolumn}
\usepackage{bm}
\def\be{\begin{equation}}
\def\ee{\end{equation}}
\def\bea{\begin{eqnarray}}
\def\eea{\end{eqnarray}}

\begin{document}
\title{A natural mechanism to induce an electric
charge into a black hole}
\author{Jos\'{e} A. de Diego$^a$, Deborah Dultzin-Hacyan$^a$,
Jes\'{u}s Galindo Trejo$^a$, Dar\'{\i}o N\'{u}\~{n}ez$^b$}
\affiliation{$^a$Instituto de Astronom\'{\i}a - UNAM, Apdo.
70-264, Ciudad Universitaria, 04510 Mexico, D.F.\\
$^b$Instituto de Ciencias Nucleares - UNAM, Apdo. 70-543, Ciudad
Universitaria, 04510 Mexico, D.F.} \email{jdo, dultzin,
galindo@astroscu.unam.mx, nunez@nuclecu.unam.mx}

\begin{abstract}
We present a natural mechanism which may induce an electric charge
into an accretting black hole in the presence of a strong, high energy
radiation field. We study this mechanism using Newtonian physics, and
we also discuss the process within the context of a Kerr Newman black
hole. Finally, we consider possible astrophysical applications in X-ray
variability and jet formation in the Active Galactic Nuclei (AGN).
\end{abstract}

\pacs{95.35.+d, 95.35.G}

\maketitle

\section{Introduction}

The standard picture for active galactic nuclei (AGN) is based on
a black hole surrounded by an accretion disk \cite{bland1978}
and, at least for radio loud objects, a jet of plasma ejected
probably through magnetohydrodynamical mechanisms (e.g.
Contopoulos \cite{conto1994}, \cite{conto1995a},
\cite{conto1995b}; Kudoh \& Shibata \cite{kudoh1997a};
Kudoh et al. \cite{kudoh1997b}, \cite{kudoh1998}).

Little importance has been given in astrophysics to mechanisms
that might produce an electric charge in a black hole, as it has
been assumed that a significant charge cannot be achieved. For
most physical situations, the electron-proton coupling in a plasma
is so strong that radiation pressure cannot break this coupling,
although it may introduce some kind of electric charge
polarization and plasma oscillations. Levich, Sunyaev \& Zeldovich
\cite{levich1972} considered first this problem in AGN,
taking into account quantum effects which introduce an extra term
for the radiation pressure. This new term is proportional to the
second power of the radiation flux and is inversely proportional
to the sixth power of the distance. In their paper, Levich et al
\cite{levich1972} showed that gravitational force may be in
fact smaller than radiation pressure with this additional term,
and that matter may escape from the nuclei of Seyfert galaxies
and quasars. However, probably due to the poor knowledge of AGN
spectral energy distribution (SED) in 1972, they considered
sources where infrared luminosity was dominant, and they argued
that: {\it`It is evident that both the nuclei and the electrons
should move simultaneously. Thus, the matter as a whole should
move away from the sources.'}

Nowadays, we know that the inner funnel and the free falling
matter into the black hole must glow with hard radiation,
although most AGN are optically deep to $\gamma$-rays.
Electron-positron pairs are created, which implies the existence
of $\gamma$ photons of at least $10^{-11}$J (i.e., $100$~MeV or
$10^{22}$Hz) interacting with soft X-ray photons. Some electrons
may be Comptonized by hard X-rays and $\gamma$-rays. An electron
Comptonized by a $10^{22}$Hz photon would acquire an energy $\sim
10^{-12}$J, i.e. two orders of magnitude larger than the electron
energy at rest. Under these circumstances, the velocity of the
electron is such that basic ideas of plasma, such as Debye
shielding, do not apply. The behaviour of a very fast particle
moving through a plasma is an important aspect of plasma physics
which has been investigated recently [e.g., Meyer-Vernet
\cite{meyer1993}, Shivamoggi \& Mulser
\cite{shiva1998}], even in the case of nuclear reactions in
stellar plasmas [Shaviv \& Shaviv \cite{shaviv1999}]. In
fact, electrons moving with velocities one order of magnitude
larger than the thermal electrons of the plasma, are practically
not affected by collisions or even magnetic fields. An
astronomical example of such fast particles are cosmic rays, both
ions and electrons, which can travel through and among the
galaxies and reach detectors on earth.

On the other hand, even though the exact solutions for charged
black holes, {\it i. e.} the Reisnner-N\"ordstrom and the Kerr
Newman metrics, have been known for more than forty years, (see
\cite{MTW}), and several analysis about their properties have
been performed (see \cite{Ruf} and references there in), those
charged black holes have always been consider of purely
theoretical interest, as long as Nature in general is neutral and
the charge to mass radio, even for an electron is so huge:
\be \frac{e}{m_e}=\frac{1.38\,10^{-34}{\rm
cm}}{6.76\,10^{-56}{\rm cm}}=2.04\,10^{21}, \ee
that even if, by some unknown process, there could be a cloud of
electrons or even protons which accreted and that, surmounting
the enormous electric repulsion, form a charged black hole, this
will not be a black hole but a naked singularity, as long as the
horizon radius, even for a non rotating black hole,
$r_H=M\left(1+\sqrt{1-(\frac{eQ}{M})^2}\right)$, turns out to be
imaginary.

In this paper, we explain how the high energetic radiation field,
generated in the inner edge of the accretion disk, can accelerate
the electrons in the outer edge of the accretion flow, allowing
them to escape from the accreting plasma. Hard X-rays and
$\gamma$-rays, can accelerate the electrons up to relativistic
velocities through Compton scattering. These charges, moving much
faster than the thermal electrons, cannot be shielded, and their
impact parameter is greatly reduced, moving almost freely through
the plasma \cite{nicho1983}. The released electrons can
eventually be incorporated into the outflow material in the inner
funnel and reach the disk corona. The efficiency of this process,
estimated from the periodicity of the X-ray variability, is very
small (approximately one out of $10^{25}$ electrons can escape).
These electrons will be supported in the corona by radiation
pressure from the disk, until the electric force from the charged
black hole overwhelms this pressure. This black hole charge is
produced by the decoupled accreted protons, and it will eventually
increase enough to reduce the accretion rate. This, in turn, will
reduce the radiation pressure until the electrons can no longer
be supported in the corona and fall to the black hole attracted
by the electric force, thus neutralizing its charge. This process
can be repeated producing low amplitude quasiperiodic or, under
some circumstances, strictly periodic variations in the X-ray
luminosity originated in the accretion disk, which may be
superposed to another component arising, for example, from the
disk corona. Moreover, we are able to compute an expression for the ratio
of the maximum charge acquired to the mass of the black hole,
and show that the black hole conserves a well defined horizon.

In section \ref{ModelPresentation} we present the model. In section \ref{KN}, we
write down the Kerr Newman solution and discuss some possible consequences of our model
within the contex of such an exact solution. In
section \ref{Observations} we apply the model to an AGN that
show X-ray periodic or quasiperiodic variability, an compare our model with other proposed models
for explaining the observed variability. And finally our conclusions are
summarized sin section \ref{Conclusions}.

\section{The model} \label{ModelPresentation}

The region between the inner edge of an accretion disk and the event horizon
of a super massive black hole contains a very high energy radiation field of X-ray and
$\gamma$-ray photons emitted at the inner edge of a thick disk. In the funnel, the radiation
is high enough to generate pairs of particles. Thus, in the
inner region, a high number of $\gamma$-ray photons is also produced, with energies of at
least $10^{-11}J$. Although high energy radiation cannot escape freely from the optically
thick accretion flow, an electron near the edge of the infalling plasma has a
probability of being scattered by an energetic photon. In this paper we
consider that the funnel is virtually empty of matter, as in
Begelman, Blandford \& Rees \cite{begel1984}.
Thus, the released electron can move freely along the funnel. However, the density
inside the funnel is an open unsolved
question. The formation of an electron-positron jet inside the funnel \cite{begel1984}
is compatible with the electron-proton decoupling as far as the $\gamma$ ray collisions
which produce electron-positron pairs imply that photons can move freely inside the
funnel, thus the funnel has an extremely low particle density. For electron-proton
plasma jets formed inside the funnel, however, optical depth
effects can become very important.

In this section we will show that, in the radiation field of very high energy
photons where the accretion flow is embedded, the pushing force of Compton
scattering on electrons and the pulling force of gravity on protons are so
intense, that they can occasionally overcome the electron-proton coupling
force in the accreting plasma. After being released from the freely falling
plasma, the electron may remain as an isolated charge and
eventually reach the disk corona, while the decoupled proton is
accreted unto the black hole inducing a positive electric charge
in the singularity. It must be noted that the main cause
underlying this process, apart from the strong gravitational
attraction of the black hole, is the high energy of the photons,
high enough to induce electron-proton decoupling, which means that
this process occurs at a microscopic rather than macroscopic
scale, as would be the case for a process induced by radiation
pressure. The efficiency of this decoupling process must
necessarily be extremely low, since a break of the global electric
coupling in a plasma is not possible. From observational data, we
will show that approximately one in $10^{25}$ electrons must be
scattered in such a way, in order to reach the disk corona and to
trigger the process. Once the scattered electron has reached the
corona, it will be decelerated by the ambient pressure and
subject to a combined pulling electric force from the charged
black hole and a pushing radiation pressure from the disk. At the
beginning of the process, the radiation pressure is very intense,
while the electric force is weak. As the positive charge in the
black hole increase by the infall of decoupled protons, the
electric repulsive force on other protons also increase and thus
the net attraction on these particles decreases, diminishing the
accretion rate and the X-ray luminosity component from the disk.
Long before the net force on the protons vanishes preventing the
accretion, the luminosity emitted in the thick part of the disk
will drop so much that the radiation pressure will no longer
support the decoupled electrons in the corona. These electrons
will fall onto the black hole at once, neutralizing it and
allowing again a higher rate of accretion, thus raising the
luminosity to its former high value. The onset of this process is
periodic, or quasiperiodic in the presence of small perturbations.

In the next subsections, we shall develop the ideas outlined above
adopting a classical or special relativity approach. These
approaches are justified because, for electrons escaping through
the inner funnel, after moving a few Schwarzschild radii,
relativistic corrections are negligible for our purpose. During
the scattering process, the electrons receive the photons with
the same energy as they had when they were emitted, with
negligible gravitational effects on the photon energy. All this
allows us to study the Compton scattering of the electrons in the
infalling plasma from energy considerations. We consider that, at
least in the edge of this plasma, the electrons can escape
without any interaction with other particles. Abramowicz \& Piran
\cite{abram1980} and Sikora \& Wilson \cite{sikora1981}
have calculated the radiative acceleration of a single particle
in an empty funnel. However, detailed descriptions of the
geodesic trajectories of the matter is beyond the scope of this
paper.

\subsection{Compton scattering on the free fall matter}

The particles in the inner accretion disk that leave the last
stable orbit fall to the black hole embedded in a piece of free
fall plasma. An electric charge in this plasma is shielded in a
distance called Debye's wavelength, which is given by:

\begin{displaymath}
  \lambda_{D} = \sqrt{\frac{\epsilon_{0} k T}{n_{e} e^{2}}}
\;.
\end{displaymath}

For a black hole of $10^{8} M_{\circ}$ ($M_{\circ}$ being the
solar mass), accreting at the Eddington
rate, the characteristic particle density near the horizon is
$10^{11}$ cm$^{-3}$ \cite{begel1984}, and the electron
temperature ranges from $10^{5}$ to $10^{7}$K. Thus, the Debye's
wavelength is of the order of $10^{-5}$ to $10^{-4}$m. This
wavelength implies a binding energy for the electron-proton pairs
in the infalling plasma, given by:

\begin{displaymath}
  E = \frac{1}{4 \pi \epsilon_{0}} \frac{e^{2}}{\lambda_{D}},
\end{displaymath}

\noindent which is of the order of $10^{-24}$ to $10^{-25}$J.

On the other hand, the infalling plasma is embedded in a highly
energetic radiation field of ultraviolet, X-ray and even $\gamma$ ray
photons, emitted from the inner edge of the disk and from the
infalling plasma itself. The maximum kinetic energy of the
`struck' electron after a Compton scattering process is given by:

\begin{equation}\label{photoenergy}
  K_{\rm max} = \frac{h \nu_{0}}{1 + 1/2\alpha}  \;,
\end{equation}

\noindent where $\nu_{0}$ is the frequency of the incident photon
and $\alpha$ is the ratio between the energies of the incident
photon and the electron in its rest frame:

\begin{displaymath}
  \alpha = \frac{h \nu_{0}}{m_{e} c^{2}} \;.
\end{displaymath}

For X-ray and $\gamma$-ray photons with $\nu_{0} \geq 10^{19}Hz$,
the electron can acquire an energy $>10^{-15}J$, which is at least
9 orders of magnitude larger than the binding energy with the
proton. Thus, the Debye's wavelength cannot be applied to such an
energetic electron. This moving charge excites electrostatic
plasma oscillations along its trajectory, trailing a train of
density oscillations of wavelength much larger than the Debye's
wavelength [e.g. Meyer-Vernet \cite{meyer1993}, Shivamoggi
\& Mulser \cite{shiva1998}].

In most cases, after a Compton scattering in a plasma, another
electron will replace the scattered one and radiation will have
either negligible effects or, if it is very intense, it may
introduce some polarization and plasma oscillations. However, near
a black hole, the infalling plasma containing the isolated proton
will be accreted and, if the scattered electron escapes from the
accretion flow, a net positive charge will be introduced into the
black hole. In fact, the electrons will be Compton scattered in
\textit{all} directions, and only a fraction of them will be able
to escape from the infall. In this context, we use the efficiency
of the process as the probability that an electron escapes from
the accretion flow through the inner funnel and reaches the disk
corona, where the radiation pressure prevents it from falling
again into the black hole. As we shall see later, this efficiency
is very low, and thus the macroscopic state of the infalling
plasma remains virtually neutral.

\subsection{Movement of a fast electron across the plasma}
\label{movement}

The plasma does not have time to respond and shield a fast
electron, i.e., an electron that is moving about one order of
magnitude faster than the thermal electrons of the plasma
[Nicholson \cite{nicho1983} and references therein]. This
implies that such a fast electron can move almost freely,
traveling a large distance before it is stopped. A measure of
this distance can be obtained starting from the collision
frequency. For an electron moving at a velocity $v$ and
interacting with another electron at a minimum distance $r_{0}$,
the impulse (change in the moment of the electron) can be
expressed as:

\begin{equation}\label{impulse}
  \Delta p \approx \frac{e^{2}}{4 \pi \epsilon_{0} r_0 v}  \;.
\end{equation}

For a relativistic electron, the initial moment $p$ is given by:

\begin{equation}\label{moment}
  p = m_{e} v \gamma \;,
\end{equation}

\noindent where $\gamma = (1-v^{2}/c^{2})^{-1/2}$ is the electron
Lorentz factor. We are interested in an increase $\Delta p \simeq
p$. Thus, substituting Equation (\ref{moment}) in Equation
(\ref{impulse}) we obtain:

\begin{equation}\label{radius}
  r_{0} \approx \frac{e^{2}}{4 \pi \epsilon_{0} m_{e} v^{2} \gamma}
\end{equation}

We can consider $r_{0}$ as the impact parameter for the
relativistic electron. Thus, the electron sweeps a volume $\pi
r_{0}^{2} v$  in a second. Then, the number of collisions per
second for a plasma of number density $n$ is:

\begin{equation}\label{freqcol}
  \nu = \frac{\pi n e^{4}}{(4 \pi \epsilon_{0})^{2} m_{e}^{2} v^{3}
  \gamma^{2}} \;.
\end{equation}

An estimate of the time that the electron moves freely through
the plasma can be obtained integrating Equation (\ref{freqcol})
over a range of impact parameters, and considering collisions
with both electrons and protons. This yields a \emph{deflection
timescale} for the electron given by [cf. Nicholson
\cite{nicho1983} for the collision frequency of a non
relativistic electron]:

\begin{equation}\label{deflection}
  t_{d} = \frac{(4 \pi \epsilon_{0})^{2} m_{e}^{2} v^{3} \gamma^{2}}
  {8 \pi n e^{4} \ln \Lambda} \;,
\end{equation}

\noindent where $\ln \Lambda$ is the Coulomb logarithm for a
plasma at a given temperature $T$ and number density $n$:

\begin{displaymath}
  \ln \Lambda \approx 10 + 3.45 \log T - 1.15 \log n \;.
\end{displaymath}

\begin{figure}[htb]
\centerline{\includegraphics[width=7cm]{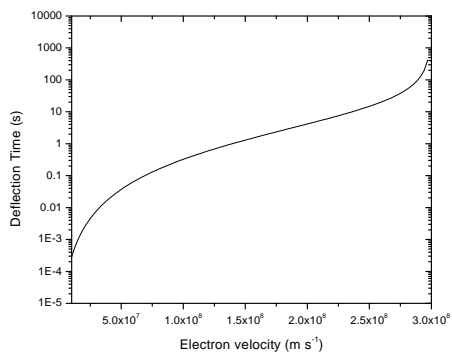}}
\caption{Deflection timescale as a function of the electron
velocity moving through a plasma of number density
$10^{11}cm^{-3}$ and temperature $10^{7}K$.}\label{figdeflec}
\end{figure}

Besides the deflection time, there are other two timescales of
interest to characterize the free path of a particle through a
plasma. In our case, the \emph{slowing-down timescale} is of the
same order as the deflection time, and the \emph{energy exchange
timescale} is an order of magnitude larger. Thus, the deflection
time is an appropriate parameter to calculate the electron free
path in our case. Figure (\ref{figdeflec}) shows the deflection
timescale for an electron moving through the accretion flow, as a
function of the electron velocity. Note that for electrons moving
faster than one third of the speed of light this timescale is
about one second.

\begin{figure}[htb]
\centerline{
\includegraphics[width=7cm]{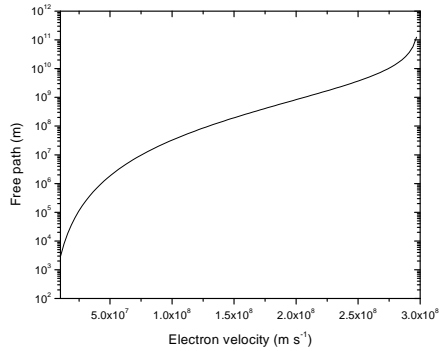}}
\caption{Free path of the electron as a function of its
velocity moving through a plasma of number density $10^{11}cm^{-3}$
and temperature $10^{7}K$.}\label{figfreepath}
\end{figure}

Thus, the free path of the electron is about
$10^{8}m$ (Fig. \ref{figfreepath}). This free path is then one
hundredth of the radius of the last stable orbit around a black
hole of $10^{7} M_{\circ}$. Thus, for Compton processes near the
outer edge of the accretion flow (about the last stable orbit),
the Comptonized electrons can reach a going on zone outside the
accreting plasma, and not be swallowed by the black hole. As for
the accreting plasma, it will sustain a small positive charge
which will be accreted unto the black hole.

The photon frequency needed to accelerate the electron through
Compton scattering can be calculated from eq. (\ref{photoenergy})
and the expression of the kinetic energy in special relativity,
resulting in:

\begin{equation}\label{nu}
  \nu = \frac{m_{e} c^{2}}{2h} \left[ (\gamma-1) +
  \sqrt{(\gamma+1)(\gamma-1)} \right] \;.
\end{equation}

\begin{figure}[htb]
\centerline{\includegraphics[width=7cm]{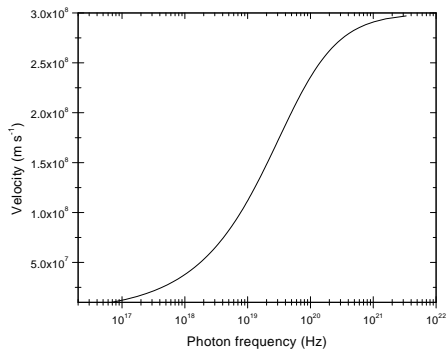}}
\caption{Comptonized electron velocity as a function of the
photon frequency.}\label{figphotfreq}
\end{figure}

Fig. (\ref{figphotfreq}) shows the velocities acquired by the
Comptonized electron as a function of the photon frequency. Note
that hard X-ray and $\gamma$-ray photons can accelerate electrons
up to one third or more of the speed of light.

\subsection{Force balance and electric discharge} \label{model}

In the previous discussion, we considered Compton scattering on
electrons as the mechanism which can induce a positive electric
charge in the black hole. However, once the decoupled electrons
have reached the disk corona, we must consider the effect of
radiation in terms of radiation pressure rather than Compton
scattering. This pressure will prevent the infall of the electrons
in the corona until the black hole reaches a critical electric
charge. For this process to take place, the black hole must not be
electrically shielded at short distances, even if the Debye
wavelength is very short. The reasons why the black hole cannot be
shielded at short distances comes from the Eddington luminosity:

\begin{equation}\label{eddington}
  L_{E} = \frac{4 \pi c G M m_{p}}{\sigma_{e}} \;,
\end{equation}

\noindent which is a measure of the maximum brightness of the disk
(although it can be surpassed for thick disks). The Eddington
luminosity depends on the mass of the particles accreted, usually
considered protons and electrons together, although only the mass
of the protons is taken in consideration. But note that if the
accreted particles where \emph{only} of the electron mass, the
luminosity would need to be three orders of magnitude lower. This
implies that single electrons cannot be accreted alone or even
efficiently shield a moderate black hole charge, as the radiation
would sweep them out. In other words, the radiation field behaves
as an electric resistance that prevents the movement of isolated
electrons towards the black hole. Then, the shielding of the
black hole would take place both in the accretion disk and in the
disk corona. But for an electron at the inner border of the
(optically) thick accretion disk, the radiation pressure from the
opposite wall overwhelms the radiation pressure from behind.
Thus, the electrons would be behind the protons. This implies
that the shielding of the black hole is performed \emph{behind}
the last stable orbit. Of course, the matter \emph{inside} the
last stable orbit cannot shield the black hole either, since there
are no stable orbits anymore.

Thus, in a classical approximation, and neglecting magnetic
effects, there are three forces acting on \textit{isolated}
particles near a black hole: gravitational, Coulombian and
radiation pressure. For protons, radiation pressure is
negligible, while for electrons, gravity force is very weak
compared to the other two. If we consider the absolute values of
these three forces, the conditions for accretion for these
particles will be:

\begin{equation}
 F_{r,e} < F_{g,e}+F_{e,e} \approx F_{e,e} \label{eq1} \;,
\end{equation}

\noindent and

\begin{equation}
  F_{g,p} > F_{r,p}+F_{e,p} \approx F_{e,p} \;, \label{eq2} \;,
\end{equation}

\noindent where the first subindex denotes gravity ($g$),
radiation ($r$) and electricity ($e$), and the second subindex
denotes proton ($p$) and electron ($e$).

All these forces are proportional to $R^{-2}$, where $R$ is the
distance from the central source, and thus a decoupled electron
will not fall again until the luminosity decreases and the charge
increase up to a certain critical limit. On the other hand, the
dependence of all these forces on $R$ permits to approach the
problem independently of $R$. Thus, eqs.~(\ref{eq1}) and
(\ref{eq2}) can be expressed as:

\begin{displaymath}
  f_{g,e} + Q f_{e} - f_{r,e} > 0 \;,
\end{displaymath}

\begin{displaymath}
  f_{g,p} - Q f_{e} - f_{r,p} > 0  \;,
\end{displaymath}

\noindent where $Q$ denotes the number of free elementary
charges. The $f$ terms are defined as:

\begin{displaymath}
 f_{g,x} = G M m_{x} \;,
\end{displaymath}
\begin{equation} \label{forces}
 f_{r,x} = L\frac{\sigma_{x}}{4 \pi c} \;,
\end{equation}
\begin{displaymath}
 f_{e} = \frac{e^{2}}{4 \pi \epsilon_{0}} \;.
\end{displaymath}

\noindent where $L$ is the luminosity, and the Thompson cross
section $\sigma_{x}$ (where $x$ is the subindex of the particle),
instead of the Klein-Nishina cross section, is used for
simplicity and because, for the soft X-ray radiation considered,
they differ only in the fourth significant digit.

The accretion rate of matter onto the black hole can be expressed
in function of the density of the material ($\rho$), a certain
surface around the black hole that the matter will cross ($S$),
and the falling velocity of this matter ($v$):

\begin{displaymath}
  \dot{M} = \rho S v \;.
\end{displaymath}

Elementary physics shows that the velocity $v$ may be expressed as
a function of the distance $R$ to the singularity and the
acceleration $a$ on the particles:

\begin{displaymath}
  v = \sqrt{2 R a} \;.
\end{displaymath}

As the luminosity is directly proportional to the accretion rate,
it will also be proportional to the square root of the
acceleration on the falling material and, therefore, to the square
root of the forces acting on the falling particles.

If we express the maximum luminosity emitted in the inner
(variable) region in terms of the Eddington luminosity ($L =
\kappa L_{E}$), and neglect the radiation pressure on the proton,
we can express the luminosity in a given moment as a function of
the forces acting on the proton:

\begin{equation}
  L = \kappa L_{E} \sqrt{\frac{f_{g,p}-Qf_{e}}{f_{g,p}}} \;.
\label{luminosity}
\end{equation}

Substituting in this expression eq.~(\ref{forces}), and replacing
$L_{E}$ for its value in function of the mass $M$ of the black
hole, Eq.(\ref{eddington}), we write the condition for
electron infall:

\begin{displaymath}
-\left[G M m_{p} - \frac{Q\,e^2}{4\pi\,\epsilon_{0}}\right] -
\kappa\,\sqrt{G M m_{p}} \sqrt{G M m_{p} -
\frac{Q\,e^2}{4\pi\,\epsilon_{0}}} + G M (m_p+m_{e}) > 0 \;.
\end{displaymath}

Solving the quadratic equation for $\sqrt{G M m_{p} - \frac
{Q\,e^2}{4\pi\,\epsilon_{0}}}$, we obtain:

\begin{displaymath}
 \sqrt{G M m_{p} - \frac{Q\,e^2}{4\pi\,\epsilon_{0}}} =
 \sqrt{G M} \frac{\sqrt{(4+\kappa^{2}) m_{p} + 4 m_{e}} -
\kappa \sqrt{m_{p}}}{2} \;,
\end{displaymath}

\noindent thus, the ratio of the critical charge to the mass of
the black hole, $\frac{eQ_{\rm crit}}{M}$, allowed before the
free electrons begin to fall to the black hole is:

\begin{equation} \label{eqQmax1}
\frac{eQ_{\rm crit}}{M} = 4 \pi \epsilon_{0}\,G\,\frac{m_p}{e}
\left[\kappa\sqrt{ \frac{\kappa^2}{4}+1+\frac{m_e}{m_p}
}-\frac{\kappa^2}{2}-\frac{m_e}{m_p} \right]\;.
\end{equation}

In cgs units, $4 \pi \epsilon_{0}=1$, $G=6.6\times
10^{-8}\frac{{\rm cm}^3}{{\rm gr}\,{\rm s}^2}$, $m_p=1.6\times
10^{-24} {\rm grs}$, $m_e=9.1\times 10^{-28} {\rm grs}$, and the
charge of the electron is $e=4.8\times 10^{-10}\frac{\sqrt{{\rm
gr\,cm}^3}}{\rm s}$, recalling that $1 {\rm Coul}= 3 \times
10^{9}\frac{\sqrt{{\rm gr\,cm}^3}}{\rm s}$. Thus,
Eq.(\ref{eqQmax1}) takes the form:
\begin{equation} \label{eqQmax2}
\frac{eQ_{\rm crit}}{M} = 2.2 \times
10^{-22}\left[\kappa\sqrt{\frac{\kappa^2}{4}+1+5.68\times
10^{-4}}-\frac{\kappa^2}{2}- 5.68\times 10^{-4}\right]\,\frac
{\sqrt{{\rm gr\,cm}^3}}{\rm s}\,\frac{1}{\rm gr}\;.
\end{equation}

Now it is easy to express this ratio in geometric units, where $c=1$, and $G=1$, thus
$1s=3\times 10^{10}{\rm cm}$, and  $1 gr=1.3 \times 10^{-28} {\rm cm}$, thus obtaining:
\begin{equation} \label{eqQmax}
\frac{eQ_{\rm crit}}{M} = 6.43 \times
10^{-17}\left[\kappa\sqrt{\frac{\kappa^2}{4}+1+5.68\times
10^{-4}}-\frac{\kappa^2}{2}- 5.68\times 10^{-4}\right],
\end{equation}
where now the charge and the mass are meassured in centimeters.
Note that for very small values of $\kappa$, the equations for
the ratio mass to charge are not very precise, due to the
simplifications done in the calculus. Thus, we expect
$\frac{eQ_{\rm crit}}{M}$ to become zero as $\kappa\rightarrow
0$, but the small term dependent of the electron mass prevents
$Q_{\rm crit}$ from reaching this limit.

\begin{figure}[htb]
\centerline{\includegraphics[width=7cm]{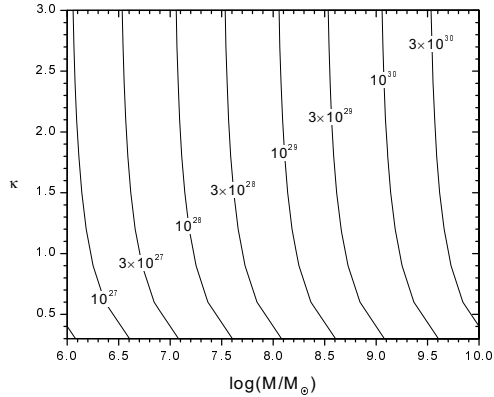}} \caption{The
contours show the dependence of $Q_{\rm crit}$ on the mass of the
black hole and the variable component of the luminosity in
Eddington's limit units $\kappa$.}\label{figcontours}
\end{figure}

To calculate the time required by the black hole to reach the
critical number of charged particles $Q_{\rm crit}$, we express
the derivative of the charge with respect to time as a function of
the accretion rate and an efficiency $\varepsilon$ for the
process. This efficiency gives the ratio between the number of
scattered electrons that will escape from the infalling plasma and
the total number of electrons that would fall to the black hole if
there were no scattering. The efficiency $\varepsilon$ may depend
on quantum effects and variables such as the binding force
between charges, and the radiation field. The evaluation of the
latter is particularly difficult since it depends on the energy
distribution and can be very inhomogeneous and anisotropic.
For this reason, we leave $\varepsilon$ as a parameter to be fixed
from the periodicity estimated from observations. Thus, the time
derivative of the charge is expressed by:

\begin{displaymath}
  \frac{dQ}{dt} = \frac{\dot{M} \varepsilon}{m_{p}} \;.
\end{displaymath}

Substituting $\dot{M} = L/\eta c^{2}$, where $\eta\approx 0.4$ is
the accretion radiating efficiency (\cite{MTW} Box 33.3), from eq.~(\ref{luminosity}),
and re-arranging we obtain:

\begin{displaymath}
\frac{dQ}{(1 - Q f_{e}/f_{g,p})^{1/2}} = \frac{\kappa \varepsilon
L_{E}}{\eta m_{p} c^{2}} dt \;.
\end{displaymath}

Integrating from zero to $Q_{\rm crit}$, we obtain the necessary
time, $\tau$, to get the critical black hole charge,

\begin{equation}
  \tau = \frac{\sigma_{e} \epsilon_{0} \eta m_{p} c}{\kappa
\varepsilon e^{2}} \left[ 2 - \left( 2\kappa^{2} - 2\kappa
\sqrt{4+\kappa^{2}} + 4 \right)^{1/2} \right] \;. \label{tau}
\end{equation}

Thus, after the lapse of time $\tau$, enough electrons will fall
onto the black hole, in order to neutralize it. In this way, the
physical initial conditions are restored and the process is
repeated. The process will be periodic or quasi\-periodic,
depending on the presence of other mechanisms that may produce
instabilities or variations of the accretion rate.

Equation~(\ref{tau}) can be approximated by:

\begin{equation}\label{tau2}
  \tau = (1.153 - 0.308 \kappa + 0.034 \kappa^2)
  \frac{10^{-20} \eta}{\varepsilon},
\end{equation}

\noindent which is a decreasing function for relevant values of
$\kappa$ (between 0 and 3).

\subsection{Amplitude of the variability}

Finally, we must consider the variation of the luminosity that
this process will produce. We define the index of variability as
the ratio between the maximum and the minimum luminosities,
$L_{\rm max}$ and $L_{\rm min}$ respectively:

\begin{equation}
  R_{L} = \frac{L_{\rm max}}{L_{\rm min}} \;. \label{deltalum}
\end{equation}

The luminosity measured by an observer on Earth may have an
underlying, constant term ($L_{c}$). This constant term can also
be expressed as a function of the Eddington luminosity:

\begin{displaymath}
  L_{c} = \chi L_{E} \;,
\end{displaymath}

Thus, eq.~(\ref{deltalum}) becomes,

\begin{displaymath}
  R_{L} = \frac{(\chi + \kappa) L_{E} }{L_{E} \{\chi + \kappa
[(f_{g,p} - Q_{\rm crit} f_{e})/f_{g,p}]^{1/2}\}} \;.
\end{displaymath}

Substituting terms and simplifying, we find that:

\begin{equation}
  R_{L} = \frac{2(\chi + \kappa)}{2\chi + \kappa
(\sqrt{4+\kappa^{2}} - \kappa)} \;. \label{deltalum2}
\end{equation}

\begin{figure}[htb]
\centerline{\includegraphics[width=7cm]{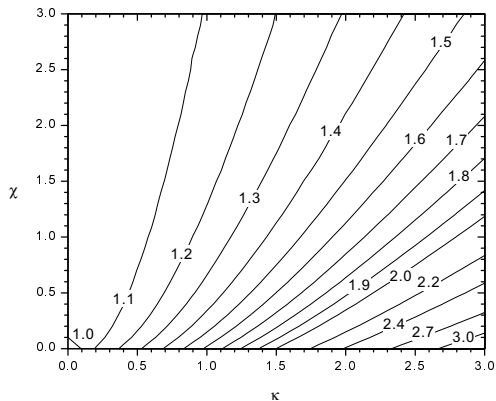}} \caption{This
figure shows the index of variability (ratio between maximum and
minimum luminosities expected from the black hole charging
process), as a function of the variable component of the
luminosity in Eddington's units $\kappa$, and the underlying
steady luminosity in the same units $\chi$. The numbers identify
isometrics of the ratio between luminosities.}\label{figindex}
\end{figure}

Fig.~(\ref{figindex}) shows a contour map for
eq.~(\ref{deltalum2}) as a function of the luminosities of the
underlying component $\chi$ and the variable component $\kappa$
in Eddington's units. In the disk model, the inner and the outer
part of the disk have similar contributions to the total
luminosity \cite{shak1973}, but the inner region is responsible
for most part of the energetic radiation. Although the Eddington
limit can be violated in some circumstances [Beloborodov
\cite{belob1998} and references therein], it must be noted
that, for a X-ray luminosity of $\kappa = 1$ and $\chi = 0$, the
ratio between the maximum and minimum luminosities $R_{L} $ has a
maximum value of approximately 1.6. This value gives an estimate
of the amplitude of the X-ray variability that our model can
account for.

It is very important to stress that, in those cases where enough
data are available to compute the bolometric luminosity, amplitude
and period of the variability, this model needs \textit{only
three} free parameters, namely the efficiency of the mass-energy
conversion for the accretion process (in this paper we use for
this efficiency $\eta=0.4$), the total X-ray and greater frequency
luminosities in Eddington units, and the efficiency $\varepsilon$
of the electron-proton decoupling process. It is also worth
stressing that the model proposed in this paper does not require
of any mechanism to enhance the variable component of the
luminosity.

\section{Kerr Newman space time} \label{KN}

It is not known an exact solution to the Einstein Maxwell equations which models the
dynamical process of accretion and charge of the black hole. Actually there is not an exact
solution for just a dynamical process of accretion in Schwarzschild! The black hole solutions are
final states of such processes. However, we think it can give us a clue of what might be happening
to consider a charged rotating black hole.

The Kerr Newman space time, which describes such a charged rotating black hole
is given by
\begin{equation}
ds^{2}=\frac{\Delta }{\Sigma }\left( dt-a\sin ^{2}\theta d\varphi
\right)^{2} -\frac{\sin^2\theta}{\Sigma}\left[ \left(
r^{2}+a^{2}\right) d\varphi
-a\,dt\right]^2-\frac{\Sigma}{\Delta}dr^2-\Sigma\,d\theta^2,
\end{equation}
where $\Delta =r^{2}-2Mr+a^2+(eQ)^2$, y $\Sigma =r^{2}+a^{2}\cos^2\theta$, $M$
being the mass of the black hole, $a$ its angular momentum per mass, and
$eQ$ its charge.

This expression is a solution to the Einstein equations,
$G_{\mu\nu}=8\pi\,T_{\mu\nu}$, with $T_{\mu\nu}$ the stress
energy tensor generated by the electromagnetic field, (in turn
generated by the electromagnetic tensor of the black hole,
$F_{\mu\nu}$), given by: \be
T_{\mu\nu}=\frac{1}{8\pi}({F_\mu}^\sigma\,F_{\sigma\nu}-\frac{1}{4}\,
g_{\mu\nu}\,F_{\alpha\beta}\,F^{\alpha\beta}).
\ee
The electromagnetic tensor $F_{\mu\nu}$ is given by:
\be F_{\mu\nu} = \left( \begin{array}{c c c c c c c} 0&-E_r&-E_\theta&0 \\
E_r&0&0&-B_\theta\\ E_\theta&0&0&B_r \\ 0&B_\theta&-B_r&0
\end{array} \right), \ee
where
\bea
E_r&=&\frac{\sqrt{2}\,eQ\,(r^2-a^2\cos^2\theta)}{\Sigma^2}, \nonumber \\
E_\theta&=&\frac{2\,\sqrt{2}\,eQ\,a^2\,r\cos\theta\,\sin\theta}{\Sigma^2}, \nonumber \\
B_r&=&\frac{2\,\sqrt{2}\,eQ\,a\,r\,(r^2+a^2)\,\cos\theta\,\sin\theta}{\Sigma^2}, \nonumber \\
B_\theta&=&\frac{\sqrt{2}\,eQ\,(r^2-a^2\cos^2\theta)\,a\,\sin^2\theta}{\Sigma^2},
\eea
and is also a solution of the Maxwell equations without
currents: ${F^{\mu\nu}}_{;\nu}=0$, and
$F_{\mu\nu;\lambda}+F_{\lambda\mu;\nu}+F_{\nu\lambda;\mu}=0$.

As mentioned in the introduction, a major concern with a model that deals with charged black holes
is that the charge must remain small enough to avoid the situation
of a \textit{naked singularity}, i.e., a black hole with no
horizon \cite{MTW}. For the Kerr-Newman metric the external horizon is given by:

\begin{displaymath}
r_+ \equiv M\left(1 + \sqrt{1-(\frac{a}{M})^2 -
(\frac{eQ}{M})^2}\right),
\end{displaymath}

\noindent where $\frac{a}{M}$ takes the canonical value
$\frac{a}{M} = 0.998$ \cite{thorne1974}. We recall the reader
that in this section we are working in the usual geometric units.
The naked singularity would be produced when
$\frac{Q}{M}>3.996\times 10^{-3}$, and from Eq. (\ref{eqQmax}),
we see that this is far from taking place as the maximum value of
the charge allowed is several order of magnitude below.

There are two invariants of the electromagnetic field, those quantities independent of the observer \cite{light}:
\bea
E^2-B^2&=&-\frac{1}{2}F_{\mu\nu}\,F^{\mu\nu}=
\frac{4\,(eQ)^2\,(r^4-6\,r^2\,a^2\cos^2\theta+a^4\cos^4\theta)}{(r^2+a^2\cos^2\theta)^4},
\nonumber \\
{\bf E}\cdot{\bf
B}&=&\frac{1}{4}\epsilon_{\mu\nu\alpha\beta}F^{\mu\nu}\,F^{\alpha\beta}
=
\frac{4\,(eQ)^2\,r\,a\cos\theta\,(r^2-a^2\cos^2)}{(r^2+a^2\cos^2\theta)^4},
\eea
from which can be constructed a third one which relates the
energy density of the electromagnetic field, ${\cal E}$, and the
magnitud of the Pointer vector of the flux, $|{\bf S}|$:
\bea
64\,\pi\,&&({\cal E}^2-|{\bf S}|^2)= (B^2-E^2)^2+4\,({\bf E}\cdot{\bf B})^2\nonumber \\
&&=\frac{16\,(eQ)^4\, (r^8 -8\,r^6\,a^2\cos^2\theta +
30\,r^4\,a^4\cos^4\theta
-8\,r^2\,a^6\cos^6\theta+a^8\cos^8\theta)}{(r^2+a^2\cos^2\theta)^8}.
\eea

These invariant quantities might prove usefull in further studies
of the mechanism present in this work within the contex of the
Kerr Newman solution. For now, we only call the reader's
attention to the fact that for small angles, which is the region
where we expect the mechanism to happen, the first invariant
takes the form (we are approximating the value $\frac{a}{M}\sim
1$, and expressing the $r$-coordinate as multiples of the
external horizon radius, $r_+$, which we are also approximating
to be equal to $M$, thus $r=nM$):
\be E^2-B^2=(\frac{eQ}{r^2})^2\,
\frac{4\,(1-\frac{6}{n^2}+\frac{1}{n^4})}{(1+\frac{1}{n^2})^4}.
\label{eq:EB} \ee
In Eq.(\ref{eq:EB}) we recognize the leading term
$(\frac{eQ}{r^2})^2$,  which is the square of the magnitude of
the classical electric field, Eq. (\ref{forces}). Notice however
that near the last stable orbit, $n\sim 3$, the relativistic
corrections for the magnitudes of the electromagnetic field are
important, tending to decrease the magnitud of the first
invariant near the horizon. Thus, as expected, our analysis is a
first approximation of the phenomena.

These facts suggest that the mechanism present would be accompained
with variable electric and magnetic fields, which in turn might trigger some
other phenomena, such as the proposed model of jet production by means of a magnetic field
switched on and off \cite{meier}.

\section{Comparison with observational data} \label{Observations}

Analysis of the X-ray light curves of AGN's have shown many
cases of short time variability. For some objects, these variations are
periodic or pseudo periodic(e.g. NGC~4151\cite{kotov1994}, 3C~273, NGC~3516,\cite{kotov1997},
NGC~5548\cite{papad1993}). We will discuss the case of the galaxy \emph{IRAS}~18325-5926 in
detail because it has a very well sampled curve.

\begin{figure}[htb]
\centerline{\includegraphics[width=7cm]{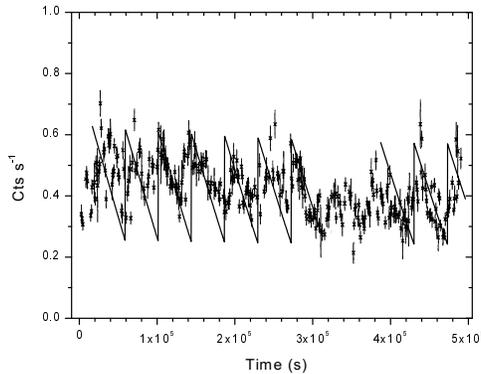}} \caption{IRAS
18325-5926 X-ray light curve from Iwasawa et al.
(\cite{iwasa1998}). Overlapped, the model fit (see main text for
explanation).}\label{figIRAS}
\end{figure}

\emph{IRAS}~18325-5926, was observed by ASCA for 5 days between 1997
March 27 and 31 \cite{iwasa1998}. Analysis of these data has
shown a periodicity of $5.8\times 10^{4}$ s in the 0.5-10 keV
band with an amplitude, obtained from the folded curve, of 15\%.
However, a visual exam of the light curve in Fig.~(\ref{figIRAS})
shows that the amplitude of the variability is rather a factor of
two larger.

In order to compare the observed light curve with the prediction
of our model, we first fit a second order polynomial to account
for long term variations. The long term variations are included
in the variable luminosity term $\kappa$ during the calculations.
Allowing a super Eddington luminosity by a factor of 2, $\kappa$
and $\chi$ are adjusted simultaneously to produce the observed
variability and luminosity in Eddington's units. We have included
in Fig.~(\ref{figIRAS}) both the observed light curve from Iwasawa
et al. \cite{iwasa1998} and the expected variability of the
X-ray luminosity from our model. We must point out that this fit
is meant only to illustrate the general expected behavior. It is
not a rigorous fit to the data in the sense of a least square or
$\chi^2$ fit. In all this process we have only set one free
parameter for this source, the luminosity in Eddington's units.
Once this parameter is selected, the others are adjusted from the
observational data.

Fig.~(\ref{figIRAS}) shows the results from the behavior predicted
by the model compared to the observations.
It is worth noticing that we are fitting the {\it
observed} light curve, rather than the folded curve. Folded
curves tend to underestimate the amplitude and structure of the
variations [cf. Kotov et al. \cite{kotov1994},
\cite{kotov1997}; Iwasawa et al. \cite{iwasa1998};
Abramowicz et al. \cite{abram1993}; Fiore et al.
\cite{fiore1992}]. Note the lag in the fitting results in
Fig.~(\ref{figIRAS}) from approximately $3\times 10^{5}$ to
$3.8\times 10^{5}$s. In this time interval, the periodical
pattern is lost. The first, small peak in this interval at
$3\times 10^{5}$s, may be still considered in phase with the
previous peaks, but with a much smaller amplitude. Then, two more
small amplitude peaks follow and the pattern is recovered at
$3.8\times 10^{5}$s, but with a different phase. This result can
be expected if the electrons fall into the central object as a
consequence of some kind of disturbance. The accumulation of
electric charge is basically an unstable process and a small
perturbation (for example, the long term variation of the source)
may produce the infall of the electronic cloud. Alternatively,
the electron distribution may be inhomogeneous. In such a case, a
Self-Organized Criticality (SOC) model [Bak, Tang, \& Wiesenfeld,
\cite{bak1988}; Mineshige, Ouchi,\& Nishimori,
\cite{mines1994}] may be applicable. The SOC model comprises
numerous reservoirs. When a critical density is reached in a
reservoir, an instability appears causing an avalanche and
emptying the reservoir. Adjacent reservoirs are coupled, and the
instability in one of them may extend to a few or many
reservoirs, resulting in a small or large flare. Small flares are
basically random, since they originate from a few reservoirs. The
reservoir model in AGN has been discussed by Begelman \& de Kool
\cite{begel1991}. It is worth noticing that for other models
(hot spots and lighthouse), the disappearance of the periodic
variability and its recovery with another phase (but the same
period and amplitude) may only be explained by invoking too many
coincidences.

The parameters used for the calculations are: The index $R_{X}$ for the X-ray variability,
$R_{X}=2$, obtained from observations. The luminosity ($L_{X+}$) at X-ray and higher
frequencies in Eddington units, $L_{X+}=2$, which is a free parameter in the
model. The observed period, taken from observation to be $\tau=43,309 s$.
The variable and constant X-ray components in
Eddington units $\kappa$, and $\chi$ respectively, are taken as
$\kappa =1.8$ and $\chi = 0.2$.These components have been computed to fit $R_{X}$ and $L_{X+}$
simultaneously, and thus they are not free parameters. The mass of the
central body, $M=2\times 10^7 M_\circ$, which is set as a
free parameter,as long as the observations do not allow to
give a definite value, so we are taking the expected value for AGN's.
Finally, the efficiency of the charging process ($\varepsilon$),
which is computed as a free parameter to fit the observed period, and
we obtain $\varepsilon=6.5\times 10^{-26}$.

And from Eq.(\ref{eqQmax}), we obtain for the maximum value of the
charge $\frac{eQ_{\rm crit}}{M}= 5.15\times 10^{-17}$, that is,
$e\,Q_{\rm crit}=2.5\times 10^9$ Coul. Notice that although it is
a huge value for the charge, the geometry of the space time
remains unaffected.

We end this section with some comments about some other
models which have been proposed to explain periodic variability in
the X-ray light curves. Perhaps the most popular is the 'hot
spots' model, which proposes that, in the innermost part of a
thick accretion disk, hot spots appear due to shocks or other
instabilities producing observable periodic variations. The
inclination of the thick disk (with respect to the observer) has
to be such that the disk itself periodically occults the spots in
the inner border: e.g. Bao \& Abramowicz \cite{bao1996}.
However, this model requires a high number
of free parameters to be adjusted to fit the {\it folded} light
curves, which make these models difficult to be testable; it requires
$3\,n +4$ free parameters, where $n$ is the
number of spots \cite{abram1993}. In contrast, the model presented in this
paper may be expressed in function of only three free parameters, namely the mass-to-energy
conversion factor $\eta$, which is deduced from the accretion
theory, the high energy luminosity in Eddington's units $L_{X+}$,
and the efficiency ($\varepsilon$) of the local electron-proton
decoupling events that push the released electrons away from the
black hole. We have applied the model to the \emph{IRAS}~18325-5926 case where X-ray
periodic or quasiperiodic low amplitude variability has been
reported, and found that the model fits reasonably well the
observational data, even for unfolded light curves. Moreover, this
is the only model that naturally explains the disappearance of the
periodicity and later reappearance with a change of phase.

\section{Conclusions} \label{Conclusions}

In this paper, we have studied a process that may induce
electrical charge in a black hole and we have explained how this
process can produce low amplitude, periodic or quasiperiodic X-ray
variability.  We have shown that high
energy radiation along with a large gravitational field generated
by a super massive black hole may be able to break a few
proton-electron couplings in the infalling plasma due to Compton
scattering. Some free electrons may reach the disk corona while
the unbound protons are accreted into the black hole. With this
mechanism, the black hole will acquire a net positive electric
charge. Released electrons will remain as isolated charges which,
once in the corona, are continuously pushed by radiation pressure.
We have shown that it is sufficient that about one out of
$10^{25-26}$ electrons is Compton scattered \textit{away} from the
infalling plasma to trigger this mechanism. The electric repulsion force between
protons and the positively charged black hole will hamper further
infall, decreasing the accretion rate and the energy output. As
the radiation diminishes, its pressure decays and the electronic
cloud in the disk corona will infall and neutralize the black
hole. The X-ray variable component of the observed luminosity,
which had been decaying gradually during the black hole charging
process, will increase dramatically in a short time interval when
the charge is neutralized, yielding a sawtooth light curve in
X-rays. The process is periodic if there are no disturbances, but
even small perturbations may provoke a significant part of the
electrons to infall. However, the periodicity will be recovered
after the disturbance ceases.

We mentioned examples of AGN's where a periodic or semi periodic variability
in the X-ray spectra has been observed, and applied our model to the case
of the galaxy \emph{IRAS}~18325-5926, where we are able to reproduce the
overall behavior of the light curve. This result, along with the fact the
model presented in this paper has only two free parameters,
(considering the efficiency of the mass-energy conversion a data),
are evidence that a simple mechanism, like the one taht we propose, may
produce the observed X-ray variability in AGN acretting black holes. Even though there is no exact solution
to the Einstein-Maxwell equations to model any accreting black hole,
comparing with the Kerr Newman solution, we notice that
this phenomenom in turn generates variability in the Electromagnetic fields which
might trigger other events as the jet production by means of a magnetic field
switched on and off.

Perhaps the main result of the present work is the fact that the model presented in this
work opens the possibility that charged black holes may
indeed exist in Nature, not isolated though, but with an accreting disk.
The mechanism to generate such charged black holes seems plausible and the
model describes general features of physical objects.

With respect to this last point, we recall that, as
mentioned in the work, there are other models proposed to explain the
observed variability in the X-ray light curve of several AGN's. Compared with
those, our model needs a much smaller number of free parameters and, as
explained, it is stable in the sense that after a perturbation, it naturally recovers
the variability in the X-ray light curve with the same characteristics. However,
we do not discard other models, but rather propose another one which can be
also occurring.

The possibility that there is a natural mechanism to induce an electric charge into
an accreting black hole surrounded by a luminous accretion disk, could be a fruitful
subject for further studies of charged black holes and its application to astrophysics.

\section*{Acknowledgments}
We dedicate this work on behalf of Professor Alberto Garc\'\i a,
a dear college of us. This work was supported by grants
PAPIIT-UNAM IN122002 and IN113002. It is a pleasure to thank
Alejandro Corichi, Shahen Hacyan, Massimo Calvani, for
enlightening discussions and important suggestions. We
acknowledge the Astronomical data Center at the NASA Goddard
Space Flight Center for the maintenance of a public database.
This research has made use of the NASA/IPAC Extragalactic
Database (NED) which is operated by the Jet Propulsion
Laboratory, California Institute of Technology, under contract
with the National Aeronautics and Space Administration.

\label{lastpage}

\end{document}